\documentclass[sigconf]{acmart}

\usepackage{tabularx,arydshln}
\usepackage{multirow}
\usepackage{soul}  
\usepackage{color} 
\usepackage{xcolor}
\usepackage{xspace}
\usepackage{xargs}
\usepackage{listings}
\usepackage{marvosym}
\usepackage{fontawesome5}
\usepackage{subcaption}
\usepackage{circledsteps}
\usepackage{enumitem}

\newcommand{\eg}{{\it e.g.,\ }}
\newcommand{\etal}{{\it et al.\ }}

\newcommand{\ie}{{\it i.e.,\ }}

\newcommand{\sys}{\textit{Reverger}}
\newcommand\myCircled[2][]{\ifmmode
\Circled[fill color=black,inner color=white,#1]{\mathsf{#2}}
\else
\Circled[fill color=black,inner color=white,#1]{\sffamily#2}
\fi
}

\definecolor{buttonbg}{RGB}{167,189,202}

\usepackage{tikz}
\newcommand\mySquared[2][]{%
  \tikz[baseline=(X.base)]%
    \node[
      fill=white,              
      rectangle,               
      #1                       
    ] (X) {\sffamily#2};       
}

\definecolor{red}{RGB}{178,34,34}

\newif\iftrack
\tracktrue 
\newcommandx{\revised}[1][]{\iftrack{\textcolor{black}{#1}}\else#1\fi}

\AtBeginDocument{%
  \providecommand\BibTeX{{%
    \normalfont B\kern-0.5em{\scshape i\kern-0.25em b}\kern-0.8em\TeX}}}

\begin{document}

\title[Scaffolding Recursive Divergence and Convergence in Story Ideation]{Scaffolding Recursive Divergence and Convergence\\ in Story Ideation}


\author{Taewook Kim}
\affiliation{%
\institution{Northwestern University}
\city{Evanston}
\state{IL}
\country{USA}
}
\email{taewook@u.northwestern.edu}

\author{Matthew Kay}
\affiliation{%
\institution{Northwestern University}
\city{Evanston}
\state{IL}
\country{USA}
}
\email{matthew.kay@u.northwestern.edu}

\author{Yuqian Sun}
\affiliation{%
\institution{Midjourney}
\city{San Francisco}
\state{CA}
\country{USA}
}
\email{ysun@midjourney.com}

\author{Melissa Roemmele}
\affiliation{%
\institution{Midjourney}
\city{San Francisco}
\state{CA}
\country{USA}
}
\email{mroemmele@midjourney.com}

\author{Max Kreminski}
\affiliation{%
\institution{Midjourney}
\city{San Francisco}
\state{CA}
\country{USA}
}
\email{mkreminski@midjourney.com}

\author{John Joon Young Chung}
\affiliation{%
\institution{Midjourney}
\city{San Francisco}
\state{CA}
\country{USA}
}
\email{jchung@midjourney.com}



\begin{abstract}

Human creative ideation involves both exploration of diverse ideas (divergence) and selective synthesis of explored ideas into coherent combinations (convergence). While processes of divergence and convergence are often interleaved and nested, existing AI-powered creativity support tools (CSTs) lack support for sophisticated orchestration of divergence and convergence. We present Reverger, an AI-powered CST that helps users ideate variations of conceptual directions for modifying a story by scaffolding flexible iteration between divergence and convergence. For divergence, our tool enables recursive exploration of alternative high-level directions for modifying a specific part of the original story. For convergence, it allows users to collect explored high-level directions and synthesize them into concrete variations. Users can then iterate between divergence and convergence until they find a satisfactory outcome. A within-subject study revealed that Reverger permitted participants to explore more unexpected and diverse high-level directions than a comparable baseline. Reverger users also felt that they had more fine-grained control and discovered more effort-worthy outcomes.

\end{abstract}




\begin{teaserfigure}
\centering
  \includegraphics[width=\textwidth]{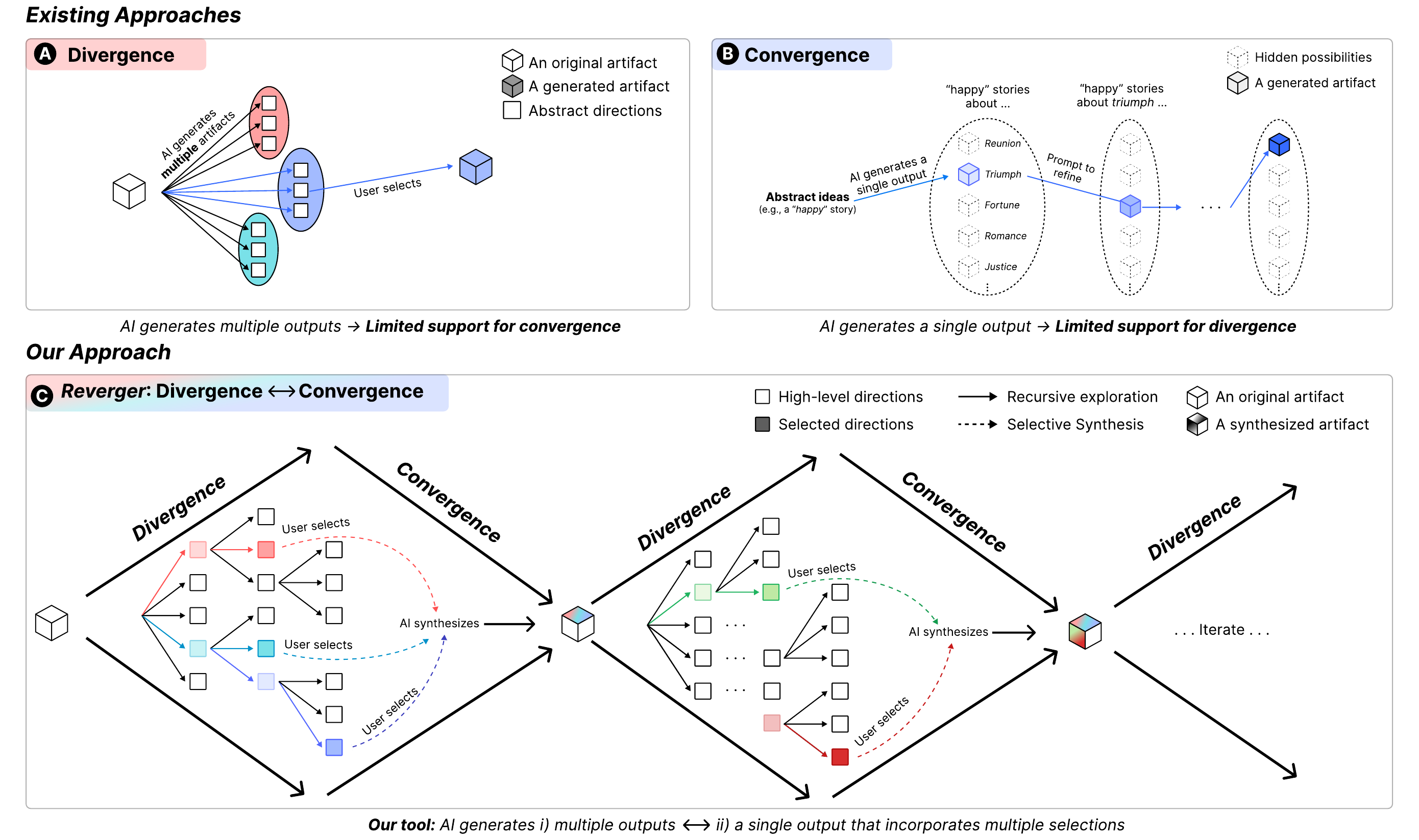}
  \caption{The conceptual illustration of our proposed interaction paradigm \myCircled{C}~and those of existing AI-powered creativity support tools (\myCircled{A} and~\myCircled{B}).~\sys{} scaffolds recursive divergence and convergence processes.
  Although existing tools (\eg Luminate) generate multiple variations (\myCircled{A} divergence), they lack support for the flexible iteration of convergence along with divergence. Some other tools (\eg ChatGPT) inadvertently promote fixation as they generate a single artifact (\myCircled{B} convergence).}
  \Description{Teaser figure description here}
  \label{fig:teaser}
\end{teaserfigure}

\maketitle

\section{Introduction}

Effective creative ideation typically involves two complementary cognitive processes: divergent thinking --- exploring diverse possibilities --- and convergent thinking --- synthesizing selected ideas into cohesive integrations~\cite{guilford1967nature, goldschmidt2016linko}. The interplay between divergence and convergence is rarely linear~\cite{goldschmidt2016linko, georgi2018enhancing}; rather, these processes are iterated, interleaved, and nested recursively across multiple layers of the creative process~\cite{mukherjee2023managing, finke1996creative, sawyer2024explaining}. Generative AI technologies (\eg LLMs) have enabled various AI-powered creativity support tools (CSTs) that demonstrate efficacy in facilitating idea generation (\eg~\cite{suh2024luminate, yuan2022wordcraft}) and implementation (\eg~\cite{toyteller2024, talebrush2022}). However, perhaps due to the generative nature of AI, existing AI-powered CSTs have focused on supporting divergence in creativity~\cite{kumar2025human}, with less attention to supporting convergence and flexible movement between these two complementary modes~\cite{gero_design_2022}.

One class of CSTs aims to support user ideation by generating diverse examples for users to consider (\myCircled{A} in~\autoref{fig:teaser}). However, these tools often primarily support lateral exploration rather than recursive exploration (\eg~\cite{gero2019metaphoria, suh2024luminate, yuan2022wordcraft, reza2024abscribe}). For instance, creative writers often wish to brainstorm variations of a story concept, exploring subcategories of high-level directions (\eg refining a high-level tonal direction like `humorous' to a more specific direction, such as slapstick, satire, or parody)~\cite{clark2018creative}. These types of tools offer limited support for this recursive exploration that characterizes natural human creative processes~\cite{mukherjee2023managing, finke1996creative, sawyer2024explaining}. In addition, real-world ideation processes involve synthesizing multiple elements concurrently, such as incorporating both `humorous' and `romantic' aspects within various `locations' in a narrative~\cite{flower1980cognition}. The interface design of existing tools (\eg~\cite{suh2024luminate}) does not fully allow users to selectively synthesize multiple ideas (\ie~\myCircled{A} in~\autoref{fig:teaser} has limited support for convergence).

Another thread of work focuses on instantiating users' vague ideas by translating abstract expressions of intent (\eg~descriptions~\cite{midjourney2025}, rough sketches~\cite{talebrush2022}) into a single concrete artifact (\myCircled{B} in~\autoref{fig:teaser}). \revised{While effective, these approaches could inadvertently promote fixation, where users select a single artifact and repeatedly iterate on it without exploring other possibilities~\cite{nigel2004expertise, jansson1991design} (\ie~\myCircled{B} in~\autoref{fig:teaser} has limited support for divergence).} Additionally, these tools take users' abstract input and generate artifacts based on AI's algorithmic interpretation instantly. This creates a significant gap between users' initial abstract intentions and the generated concrete artifacts, which is filled arbitrarily by the AI. The interfaces of existing tools do not provide sufficient spaces for users to gradually refine and compound their intentions throughout the creative process, limiting users' control in shaping the outcome.

Therefore, to support creative ideation effectively, an integrated CST is needed that enables the following: (i) interleaved and nested iteration between divergence and convergence until users discover a desirable outcome; (ii) recursive exploration of diverse possibilities across multiple layers; (iii) selection of multiple abstract directions to instantiate concrete artifacts that reflect user's evolving creative intentions (see~\myCircled{C} in~\autoref{fig:teaser} for conceptual illustration). We introduce \textit{\sys}, an AI-powered CST that scaffolds flexible and recursive iteration of divergence and convergence in story ideation (see~\autoref{fig:system}). For divergence, users can highlight a specific passage in a story and click the \faSearch~button (D.1) to generate eight potential high-level variation directions. Users can then recursively explore more specific sub-directions (\eg~refining \textit{resilience} into \textit{perseverance}, \textit{optimism}, or \textit{patience}) by clicking the \faPlusCircle~button (D.2), continuing this exploration until finding desired directions. For convergence, Reverger enables users to selectively collect and synthesize these explored directions through two features: users can select multiple promising directions by checking~\faCheckSquare~button (C.1) and then click the~\Rewind~button (C.2) to generate coherent story variations that integrate these selected directions. Through this interleaved and nested divergence and convergence workflow, users can explore variations of story concepts and select a satisfactory outcome to replace the original highlighted passage (with~\Rewind~(C.3)).

We recruited 16 participants (7 experts and 9 hobbyists) from an online freelancing platform to evaluate \textit{\sys{}}. We conducted a within-subject study to compare our tool with a baseline tool. We found that our system, \textit{\sys{}}, enabled our participants to explore more variations and discover better effort-worthy outcomes by exploring fewer variations. We also observed diverse usage patterns among participants; some utilized our tool to refine more localized text phrases or expressions, while others used it for broader conceptual brainstorming and significant story mutations. Interestingly, many participants expressed a strong sense of ownership over their outcomes, which could be attributed to our interface design offering spaces for users to infuse and refine their intention with greater control over LLM outputs. Based on these findings, we reflect on our approach and discuss several future directions: scaffolding divergence and convergence for creative ideation (\autoref{dis:proximal}) for a proximal future, preserving human agency within human-AI interaction for a distal future (\autoref{dis:agencyhai}), and limitations (\autoref{dis:limit}).

\section{Related Work}
\label{rw}
\subsection{Cognitive Processes of Human Creativity}

\subsubsection{Divergent and Convergent Thinking}
\label{rw:div-con}

Human creativity involves two cognitive processes: \textit{divergent} and \textit{convergent} thinking~\cite{guilford1967nature, goldschmidt2016linko}. Divergent thinking generates a broad range of possible ideas without immediate judgment (lateral exploration), while convergent thinking is responsible for evaluating these ideas to determine the most desirable ones (vertical exploration)~\cite{baer2014creativity, guilford1967nature, runco1991eval, goel2014creative, Brophy2001comparing}. The evaluative process of convergent thinking is not simply responsible for selecting ideas: it can also entail a more sophisticated process of synthesizing elements to create new ideas. More precisely, with convergent thinking, people can analyze generated ideas more deeply, extract valuable elements from them, and synthesize those extracted elements to create new integrated solutions~\cite{lubart2016creativity, willemsen2023role}.

The interplay between divergent and convergent thinking is recursive and concurrent~\cite{goldschmidt2016linko, georgi2018enhancing}. Throughout the creative process, people often flexibly alternate between generating ideas and evaluating them~\cite{mukherjee2023managing}. For instance, the process of narrowing options and synthesizing elements during convergent thinking could spark new ideas that may trigger subsequent phases of divergent thinking. This recursive pattern might continue until a satisfactory solution emerges~\cite{finke1996creative, sawyer2024explaining}, allowing people to refine their ideas through multiple iterations of divergent and convergent thinking~\cite{csikszentmihalyi2009creativity}.

Despite the crucial role of convergent thinking in human creativity~\cite{runco1991eval, runco2010creativity}, existing creativity support tools in the HCI community largely focus on empowering idea generation (divergent thinking) (\eg~\cite{suh2024luminate, gero2022sparks, kim2023meta, suh2022code, gero2019metaphoria}) with limited support for the evaluative and synthesizing processes (convergent thinking)~\cite{gero_design_2022}. Therefore, given the nature of the interplay between divergent and convergent thinking, we designed our tool to support the flexible and recursive iteration between these processes: generating ideas and refining them through evaluation and synthesis.

\subsubsection{Abstraction and Instantiation}
\label{rw:ab-in}
Abstraction and instantiation facilitate the flexible iteration process of divergent and convergent thinking. In divergent thinking, abstraction enables generating diverse ideas across mutually exclusive categories, promoting flexibility and originality in ideation~\cite{preiss2020relationship}. Indeed, people often generate abstract and category-level ideas without fully fleshing them out in detail. For example, design practitioners typically play with low-fidelity prototypes and rough sketches to explore multiple conceptual paths before committing to detailed development~\cite{savvas2024promptinfuser, lam2023model}. Similarly, in creative writing, authors often begin with categorical themes like `romance' or `dystopian' when ideating and developing stories, characters, or plot elements~\cite{alnajm2015main, park2024character, tuhin2024art}.

On the other hand, instantiation --- the process of translating abstract ideas into concrete and specific examples --- supports the evaluative process of convergent thinking~\cite{finke1996creative}. As explained in~\autoref{rw:div-con}, people flesh out several promising abstract ideas to assess their viability. This instantiation enables more thorough evaluation across multiple dimensions such as novelty, feasibility, and relevance~\cite{dean2006identifying}. Design practitioners often develop high-fidelity prototypes to further evaluate their ideas in more detailed aspects, such as feasibility that would not be foreseen during the ideation process~\cite{dow2011parallel, lim2008anatomy}. In addition, the instantiation of abstract ideas enables people to envision how the synthesis of multiple ideas could look in tangible examples~\cite{fauconnier2002way, beaty2014roles}. For example, a creative writer may want to select both `romance' and `dystopian' themes. They might then draft (instantiate) an example story of romantic relationships within a dystopian society.

To sum up, abstraction and instantiation serve as complementary cognitive processes in human creativity. We integrated these workflows into our tool design to support: (1) idea generation at the abstract, conceptual level to facilitate flexible exploration across diverse categories (divergence), and (2) the instantiation of desirable conceptual ideas into concrete examples for idea evaluation and refinement (convergence).

\subsection{Creativity Support Tools with LLMs}

Previous HCI research has categorized creativity support tools (CSTs) that facilitate three different processes: supporting ideation, implementation, and evaluation~\cite{chung2021inter, jonas2019mapping}. With recent advances in generative AI models (\eg~LLMs), extensive work has introduced AI-powered CSTs that demonstrate enhanced capabilities supporting these processes~\cite{kumar2025human}. Specifically, LLMs can generate diverse examples to support divergent thinking for inspiration~\cite{gero2022sparks, kim2023meta, yuan2022wordcraft, suh2024luminate, cai2023design}, translate abstract ideas into concrete artifacts~\cite{talebrush2022, chung2024patchview, toyteller2024}, and provide feedback for evaluation~\cite{tuhin2024art}.

\subsubsection{Supporting Ideation}
One major thread of LLM-powered CSTs focuses on supporting divergent thinking by generating a wide range of ideas. These tools inspire new creative directions while reducing costs (\ie time and effort) for creators to generate them. For example, Metaphorian~\cite{kim2023meta} and Wordcraft~\cite{yuan2022wordcraft} generate multiple examples to support ideation and story drafting. Building on this approach, Suh~\etal introduced Luminate~\cite{suh2024luminate}, a framework that provides structured generation of diverse high-level directions and corresponding example stories to help users avoid fixation and explore alternative possibilities. While these tools could support divergent thinking, they present several limitations. As mentioned earlier in~\autoref{rw:div-con}, human creativity requires flexible and recursive iteration between divergent and convergent thinking processes. However, existing work (\eg~\cite{suh2024luminate, reza2024abscribe}) has not assessed recursive exploration of high-level directions, limiting the iterative refinement process that creators repeat until they discover satisfactory outcomes~\cite{finke1996creative, sawyer2024explaining}. Additionally, the interfaces of prior systems support only the two-dimensional synthesis of high-level directions (\eg~\cite{suh2024luminate}), whereas users might want to combine multiple creative directions simultaneously (\eg humorous, romantic, dystopian > a \textit{humorous} story of a \textit{romantic} couple in a \textit{dystopian} society). Given these limitations, we designed our tool to afford: i) recursive exploration of abstract directions, ii) flexible iteration between divergent and convergent thinking processes, and iii) synthesis of multiple abstract directions according to users' creative needs.

\subsubsection{Supporting Implementation}
Another thread of CSTs focuses on translating abstract ideas into concrete artifacts. These tools enable users to create concrete artifacts by describing their intentions (\eg a simple prompt~\cite{chatgpt, midjourney2025}, a visual sketch~\cite{talebrush2022}) without requiring extensive effort or expertise. For instance, Chung~\etal introduced Talebrush~\cite{talebrush2022}, which generates a complete story based on the user's line-drawing representations of the intended narrative, and Toyteller~\cite{toyteller2024}, which generates a narrative scene of characters based on the user's manipulation of symbolic icons. While effective, these tools present a fundamental challenge. They typically generate a single instance of a high-level direction that could potentially yield countless valid representations~\cite{suh2024luminate}. It is not always clear to the user how the AI tools choose one particular output to present among a large set of outputs spanning the valid space. The seemingly arbitrary and automatic instantiation of user intention can diminish authorial control, which could compromise the sense of ownership~\cite{kim2024authors, kim2023meta} and creative agency~\cite{zhang2024VR, anderson2024homo}. Such interfaces provide limited spaces for users to express and refine their fine-grained intentions. Therefore, we propose an interface design that enables users to infuse and iteratively refine their intent, thereby bridging the gap between abstract user intentions and AI's arbitrary outputs while having meaningful creative control.

\begin{figure*}[h!]
  \includegraphics[width=0.9\textwidth]{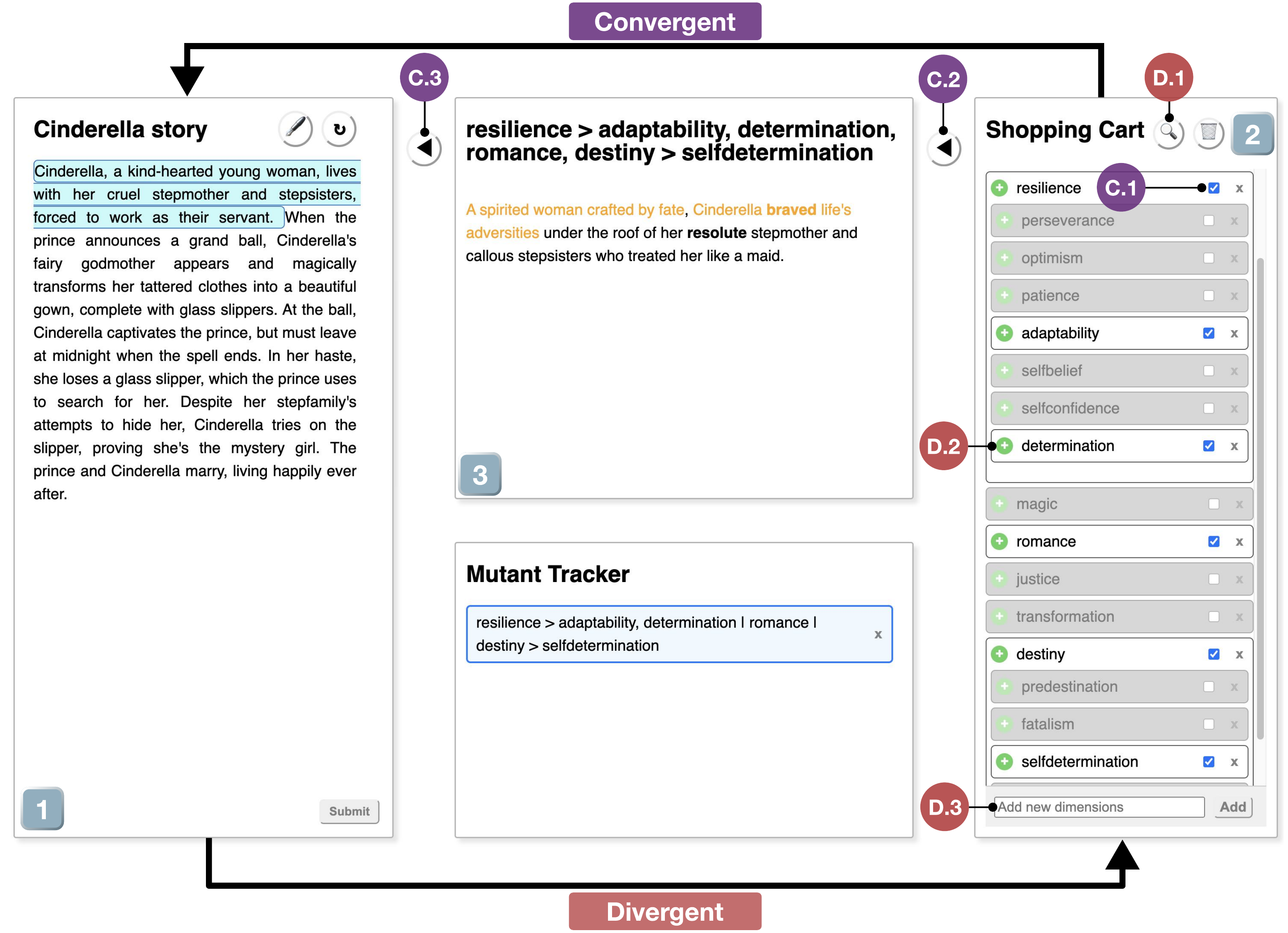}
  \caption{The interface of \textit{Reverger}. 
  While users can directly edit texts in the story box (\mySquared[fill=buttonbg,text=white]{1}), they can highlight specific passages to explore variations. Users can click~\faSearch~(D.1) to generate the breadth of high-level directions of possible variations in the shopping cart (\mySquared[fill=buttonbg,text=white]{2}). They may generate sub-directions recursively by clicking \faPlusCircle~(D.2). Next, users can selectly collect multiple directions by marking \faCheckSquare~(C.1), and generate synthesized variation examples by clicking \Rewind~(C.2). After reviewing generated variations in the example box (\mySquared[fill=buttonbg,text=white]{3}), users can replace the highlighted part with the outcome by clicking~\Rewind~(C.3).}
  \label{fig:system}
\end{figure*}

\subsubsection{Supporting Evaluation}
A third category of LLM-powered CSTs focuses on supporting the evaluation process by generating feedback. Many studies propose using LLMs to generate feedback for potential benefits such as easy reproduction~\cite{chiang-lee-2023-large} and diverse perspectives~\cite{hayati-etal-2024-far}. However, recent research demonstrates that LLM-generated evaluations are often unreliable and do not align with human assessments~\cite{si2025can}. In human creative processes, evaluation inherently demands significant cognitive resources and attention~\cite{Gabora2010revenge}, which naturally incorporates individual perspectives and preferences. So to speak, evaluation is the activity where a `weighted bias' (\ie~personal values and expertise) can be reflected in artifacts. This subjective and unique aspect of artifacts represents an essential expression of human agency in the creative process~\cite{kim2024authors}. Outsourcing evaluative judgments to LLMs could compromise this human agency, allowing algorithmic biases (\eg~\cite{dudy2025unequal, anderson2024homo}) to supersede human discernment. While the machine learning community increasingly focuses on aligning AI with human assessment (\eg~\cite{shaikh2025aligning}), we approach this issue from an HCI perspective by designing an interface that leverages LLMs as supportive tools rather than replacements for human judgment. Based on this design philosophy, we designed the interface of our tool.

\section{Design Principles}
Reflecting on the discussion in~\autoref{rw}, we summarized several design principles that we aimed to represent through our tool.

\begin{itemize}
    \item \textbf{Divergence $\leftrightarrows$ convergence}: supporting interleaved and nested iteration processes between divergence and convergence until users discover a satisfactory outcome. 
    \item \textbf{Divergence through abstraction}: supporting recursive exploration of diverse high-level directions across multiple layers until users discover a satisfactory direction.
    \item \textbf{Convergence through instantiation}: supporting selective collection and instantiation of multiple high-level directions into a cohesive artifact to refine their intentions.
\end{itemize}

\section{Reverger: Scaffolding Divergence and Convergence in Story Ideation}

With our design principles in mind, we built \sys{}, an LLM-powered tool that scaffolds creative ideation to modify the concepts of an original story in various directions. Our tool can be used in two possible cases: when users want to modify the original story with vague intentions, and when users want to transform the format of an original story (\eg literary adaptation~\cite{krasilovsky2017great, desmond2015adaptation}). \autoref{fig:system} shows the overview of \textit{Reverger}. Our tool allows users to highlight specific parts of an original story (\mySquared[fill=buttonbg,text=white]{1} in~\autoref{fig:system}), to recursively explore various directions for story modifications (\mySquared[fill=buttonbg,text=white]{2} in~\autoref{fig:system}), and synthesize multiple directions into cohesive artifacts (\mySquared[fill=buttonbg,text=white]{3} in~\autoref{fig:system}). We detailed the workflow sequence and design of \sys{} below.

\subsection{Interface and Features}
\label{sys:features}

\textit{Reverger} has four panels: \textit{story board}~\mySquared[fill=buttonbg,text=white]{1}, \textit{shopping cart}~\mySquared[fill=buttonbg,text=white]{2}, \textit{example box}~\mySquared[fill=buttonbg,text=white]{3}, and \textit{mutant tracker}; and two complementary components: features supporting \textit{divergence} (D.1-D.3) and \textit{convergence} (C.1-C.3) (see~\autoref{fig:system}). Specifically, the divergence-supporting features are designed to enable creative exploration:
\begin{itemize}
    \item \faSearch~(D.1) generates multiple abstract directions for modifying the original story.
    \item \faPlusCircle~(D.2) generates multiple sub-directions to enable recursive exploration of abstract directions.
    \item $[$Add $]$~(D.3) allows users to manually add custom abstract directions for flexibility.
\end{itemize}
The convergence-supporting features help refine and implement ideas with more user control:
\begin{itemize}
    \item \faCheckSquare~(C.1) allows users to select multiple directions to infuse their intentions into instantiating a synthesized variation.
    \item \Rewind~(C.2) generates concrete examples that coherently synthesize user-selected abstract directions for user assessment.
    \item \Rewind~(C.3) replaces the highlighted part with variation outcomes to let users update the original draft.
\end{itemize}
Our interface design supports a recursive and iterative creative story ideation workflow. Below, we present how \sys{} incorporates these features into the divergence and convergence workflow.

\subsubsection{Features Supporting Divergence Workflow:~\mySquared[fill=buttonbg,text=white]{1} $\rightarrow$~\mySquared[fill=buttonbg,text=white]{2}}
\label{sys:divergence}

In the divergence workflow, users can first highlight specific parts of the original draft in the \textit{story board}~\mySquared[fill=buttonbg,text=white]{1}, allowing them to work on the draft in smaller segments, refining each piece individually~\cite{yuan2022wordcraft, reza2024abscribe}. The system also includes a \faSearch~button (D.1) in the \textit{shopping cart}~\mySquared[fill=buttonbg,text=white]{2} that generates eight contextually relevant and semantically disjoint high-level directions to modify the story concept. We used eight to show a sufficient number of high-level directions without overloading users. Furthermore, the \faPlusCircle~button (D.2) enables users to recursively explore sub-directions of each direction, offering more granular and nuanced options to discover more variations. In case the system does not show any high-level directions users are looking for, our interface provides the [add] button (D.3) to let users manually type and add custom directions.

\subsubsection{Features Supporting Convergence Workflow:~\mySquared[fill=buttonbg,text=white]{2} $\rightarrow$~\mySquared[fill=buttonbg,text=white]{3} $\rightarrow$~\mySquared[fill=buttonbg,text=white]{1}}
 
After exploring various abstract directions in the in the \textit{shopping cart}~\mySquared[fill=buttonbg,text=white]{2}, users can select multiple explored directions by marking \faCheckSquare~(C.1). Then users can click the \Rewind~button (C.2), which prompts the LLM to instantiate a concrete example that integrates all user-selected directions into a coherent narrative piece.
For instance, if a user selects `humorous > slapstick' and `setting > location', the LLM will generate a modified story incorporating both elements (see~\autoref{tab:recursive} for more examples). The machine-instantiated variation examples would appear in the example box~\mySquared[fill=buttonbg,text=white]{3}, while a corresponding label would be added to the Mutant Tracker. Each subsequent click of \Rewind~(C.2) generates a new example synthesizing the same set of selected abstract directions. This lets users explore multiple concrete variations. Users can access previous examples by clicking on the labels in the Mutant Tracker for easy comparison and iteration. Finally, once users find a satisfactory variation, they can hit \Rewind~(C.3) to update the highlighted parts in the storyboard~\mySquared[fill=buttonbg,text=white]{1} with the chosen variation from box~\mySquared[fill=buttonbg,text=white]{3}. This functionality enables flexible integration of preferred variations into the main story.

\subsection{System Implementation}
\label{sys:implementation}

\begin{figure*}[h!]
  \includegraphics[width=\textwidth]{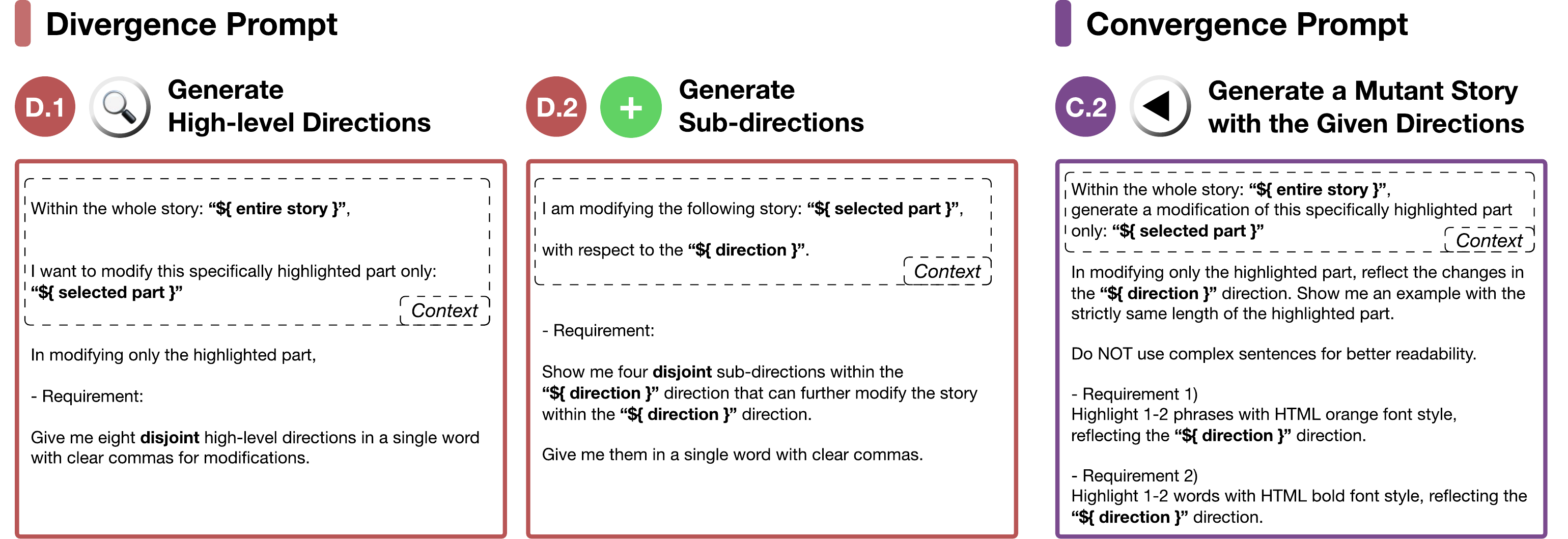}
  \caption{LLM prompt design. When users click buttons (D.1, D.2, and C.2), a new Claude API call is sent with the corresponding prompts. To ensure the contextual relevance, all prompts include the \texttt{context} of the whole story along with the selected passage.}
  \label{fig:prompt}
\end{figure*}

\textit{Reverger} is a web application with a React.js front-end and an Express.js back-end. The generative features are powered by Anthropic's \texttt{claude-3-sonnet-20240229}~\cite{claude3} model API of default settings. Below, we explain our LLM prompts for each feature.

\subsubsection{LLM Prompt to Generate High-level Directions (Divergence)}

We designed the prompt for \faSearch~(D.1 in \autoref{fig:prompt}) to generate eight distinct high-level directions tailored to the highlighted part (\$\{selected part \}) while still considering the overall story context (\$\{ entire story \}). We crafted the prompt of \faPlusCircle~(D.2 in \autoref{fig:prompt}) to generate four sub-directions for any given directions (\$\{ direction \}) recursively. To ensure relevance, in this prompt, we also provided the context of the currently selected part (\$\{ selected part \} in~\autoref{fig:prompt}) of the story.

\subsubsection{LLM Prompt to Generate Mutant Stories (Convergence)}

Users first need to select multiple abstract directions by marking~\faCheckSquare~(C.1). Then our tool can generate instances that synthesize all marked directions into a cohesive artifact. These marked directions are concatenated and embedded in the prompt of \Rewind~(C.2). To achieve the functionality of \Rewind~(C.2), we designed the LLM prompts labeled C.2 in \autoref{fig:prompt}. It first provides the entire story context (\$\{ entire story \}), followed by the selected parts to explore variations of (\$\{ selected part \} in~\autoref{fig:prompt}). Then, it provides high-level directions (\$\{ direction \} in~\autoref{fig:prompt}) as a prompt variable to instruct the LLM to integrate these multiple directions when generating mutant stories.

In our heuristic attempts with LLMs, we noticed several issues in the generated output. The story variations often became overly complex, exhibiting issues such as excessively complex language and significantly longer outcomes than the original (also noted by~\cite{tuhin2024art}). To minimize these issues, we added several constraints for the complexity of sentences and the length of the outcomes (see C.2 in~\autoref{fig:prompt}). Additionally, we designed LLM prompts to indicate specific words and phrases that evidently reflect the selected directions in the generated mutant story (C.2 - Requirements 1 and 2 in~\autoref{fig:prompt}). This addition could help users easily identify and understand changes based on their selected directions.

\section{User Study}

Following the evaluation approach suggested by Li~\etal~\cite{li2023beyond}, we conducted an exploratory study focusing on how well \sys{} supports divergence-convergence practice in story ideation. Specifically, we investigated the following research questions:
\begin{itemize}
    \item \textbf{RQ1:} Does our tool help writers iterate convergence and divergence to discover desiderata in creative story ideation?
    \item \textbf{RQ2:} How would creative writers incorporate our tool into their creative writing workflow?
\end{itemize}
To answer these questions, we compared \sys{} with a baseline tool across two different initial stories to obtain feedback from potential user groups (\ie creative writers). We employed the Latin Square design~\cite{neil2001international} (see~\autoref{tab:conditions} for details) to counterbalance the order of tools and stories. Each participant completed two sessions, using both tools in different stories (\eg session 1: our system with story A; session 2: baseline with story B). Below, we describe the baseline setup (\autoref{study:baseline}) and the stories we used (\autoref{study:material}).

\subsection{Baseline Setup}
\label{study:baseline}

For the baseline tool, we disabled two key features of \sys{}: recursive exploration of high-level directions to modify the original story~\faPlusCircle~(D.2) (divergence) and selective collection of multiple directions to synthesize them into coherent story variations~\faCheckSquare~(C.1) (convergence). With the baseline tool, users could still highlight specific parts of the original draft and explore eight high-level directions by clicking~\faSearch~(D.1). However, the exploration of high-level directions was limited to one layer with their sub-directions without the recursive nature. In addition, users can select only one high-level direction by checking~\faCheckSquare~(C.1), not multiple ones at once. This baseline setup was designed to represent the limitations of recent existing tools (\eg~\cite{suh2024luminate, reza2024abscribe, yuan2022wordcraft}) and to evaluate whether our system's divergence-convergence support features would be beneficial for the story ideation workflow.

\begin{table*}[h!]
\caption{Participant backgrounds and the Latin-Square design (2 $\times$ 2) for task assignment for each participant. *We specified their experiences with LLMs --- any prior experiences with LLMs and the creative writing experience with LLMs.}
\label{tab:conditions}
\centering
\resizebox{\textwidth}{!}{%
\begin{tabular}{l l l l l l c l l l l }
\toprule
    & & & & & & & \multicolumn{2}{c}{\textit{Session 1}} & \multicolumn{2}{c}{\textit{Session 2}} \\ \cmidrule(lr){8-9} \cmidrule(lr){10-11}
    
    & Age & Gender & Profession & Expertise & Year & Exp. w/LLMs* & System & Story & System & Story \\
    \midrule
    
    P1 & 36 & Female & Legal Assistant & Hobbyist & 15 & Yes | No & Reverger & A & Baseline & B \\ 
    P2 & 51 & Male & Freelancer & Hobbyist & 2 & Yes | No & Reverger & A & Baseline & B \\ 
    P3 & 24 & Male & Freelancer & Expert & 2 & Yes | No & Reverger & A & Baseline & B \\ 
    P4 & 24 & Male & Product Manager & Hobbyist & 4 & Yes | Yes & Reverger & A & Baseline & B \\ 
    \hdashline
    P5 & 21 & Female & College Student  & Hobbyist & 3 & No | No & Baseline & A & Reverger & B \\ 
    P6 & 29 & Female & Freelance Editor & Expert & 5 & Yes | Yes & Baseline & A & Reverger & B \\ 
    P7 & 24 & Female & Freelancer & Expert & 4 & Yes | No & Baseline & A & Reverger & B \\ 
    P8 & 64 & Male & Freelancer (retired) & Hobbyist & 40+ & Yes | Yes & Baseline & A & Reverger & B \\ 
    
    \hdashline
    P9 & 39 & Female & Marketer & Expert & 18 & Yes | Yes & Reverger & B & Baseline & A \\ 
    P10 & 40 & Female & Freelancer & Expert & 6 & Yes | Yes & Reverger & B & Baseline & A \\ 
    P11 & 34 & Male & Software Engineer & Hobbyist & 4 & Yes | No & Reverger & B & Baseline & A \\ 
    P12 & 34 & Female & HR Specialist  & Hobbyist & 25 & Yes | Yes & Reverger & B & Baseline & A \\ 
    
    \hdashline
    P13 & 41 & Female & Service Advisor & Hobbyist & 6 & Yes | No & Baseline & B & Reverger & A \\ 
    P14 & 74 & Male & Freelancer (retired) & Expert & 40+ & Yes | Yes & Baseline & B & Reverger & A \\ 
    P15 & 34 & Female & Journalist & Expert & 5 & Yes | No & Baseline & B & Reverger & A \\ 
    P16 & 31 & Female & Consultant & Hobbyist & 15 & Yes | No & Baseline & B & Reverger & A \\ 
    
\bottomrule
\end{tabular}
}
\end{table*}

\subsection{Story Materials}
\label{study:material}
We chose two widely recognized literary stories: (A) \textit{Cinderella story} and (B) \textit{Alice in Wonderland}. By providing universally familiar stories as materials, we tried to ensure that participants focus on the story ideation task without being distracted by understanding the creative stories provided~\cite{kim2024authors}. In addition, the choice of these stories addresses practical considerations, as they are both in the public domain and free from copyright restrictions. Note that we generated a one-paragraph summary of these two stories through \texttt{claude-3-sonnet-20240229}~\cite{claude3}. We found that some generated summaries overly omit important scenes of the story (\eg failing to describe how Cinderella meets the Prince). Hence, we selected one that hits such core aspects of the story.

\subsection{Participants}
\label{study:participants}

We intended to recruit participants with a wide range of expertise (\ie hobbyists and experts) in creative writing. This was meant to gain a comprehensive understanding through a spectrum of participants using our tool (\eg~\cite{wang2024lave, talebrush2022}), not to compare between distinct groups. We recruited 16 participants (10 female and six male; 9 experts and 7 hobbyists) from Upwork\footnote{\url{https://www.upwork.com/}}, an online freelancing platform. Our participant pool includes two experts with extensive experience over 18 years (P9) and 40 years (P14), five experts with relatively limited experience around five years or less (P3, P6, P7, P10, and P15), and nine hobbyists with varying levels of experience (ranging from two years to 40+ years). The participants self-identified as hobbyists or experts. All experts published their work, while none of the hobbyists did so. We tabulated details and task assignments of our participants in~\autoref{tab:conditions}.

We asked participants about their experiences with LLMs, particularly about their writing activities. All participants had prior experiences with LLMs (\eg ChatGPT~\cite{chatgpt}, Gemini~\cite{gemini}), except P5. Seven participants have been using LLMs for their various writing tasks such as brainstorming (\eg P2, P14) and editing email (\eg P12, P15). Each participant received 35 USD as compensation for about a 70-minute study upon completing all tasks (both sessions 1 and 2) and post-study interviews (see~\autoref{method:procedure} for details).

\subsection{Procedure}
\label{method:procedure}
We conducted a 1:1 synchronous online study via Zoom. Upon participants' arrival, we gave an overview of the study procedure. After obtaining participants' consent, we started audio and screen recordings for data analysis. Prior to introducing any tools, we conducted a brief pre-study interview (lasting less than five minutes) to gather background information, including participants' occupations, prior experiences with LLMs, and their motivations for creative writing. All participants went through two tutorial sessions; one with the baseline and another with \sys{}. To mitigate potential ordering effects, we counterbalanced the sequence of tools across participants (see~\autoref{tab:conditions}).

\subsubsection{Tutorial (10 mins)}

We gave participants a link to access the system remotely. We asked them to share their screens and grant us remote control for our tutorial demonstration. We followed a structured script during the tutorial to ensure consistency in the quality and quantity of information provided to all participants. We introduced our tool as an AI-powered tool designed to assist writers in their story ideation practices. We explained the key features of two tools: divergence supporting features (D.1-3) and convergence supporting features (C.1-3) for \sys{}, and restricted features for the baseline (see~\autoref{fig:system}). We demonstrated interaction scenarios, showing how users could recursively explore the breadth and depth of possible directions to modify the original story, and generate the mutant story pieces through the selective collection of multiple directions flexibly. Following the demonstration, participants were allowed to play with the tool freely. Throughout this tutorial period, we encouraged participants to ask any questions. It took about 10 minutes for the baseline and our system, respectively.

\subsubsection{Tasks (10 mins)}
\label{method:task}

After the tutorial, participants began a story ideation task using the provided tool (either the baseline or \sys{}). Participants were given free rein to modify the story as they wanted to pursue. Researchers observed the participants' shared screens in real-time and addressed any questions or issues that arose during the task sessions. Following past work~\cite{yuan2022wordcraft}, we gave participants 10 minutes to complete the task. However, we allowed them to continue if they insisted (\eg P7, P15). After the completion of each task (with the baseline and \sys{}), we asked participants to fill out the NASA Task Load Index (TLX)~\cite{HART1988} and Creativity Support Index (CSI)~\cite{cherry2014csi} survey questionnaires to measure their perceived cognitive load and the tool's creativity support, respectively. Finally, we asked participants to share specific examples of what they liked and disliked about the tool to capture their immediate impressions.

\begin{figure*}[h!]
    \centering
    \begin{subfigure}[b]{0.48\textwidth}
        \centering
        \includegraphics[width=\textwidth]{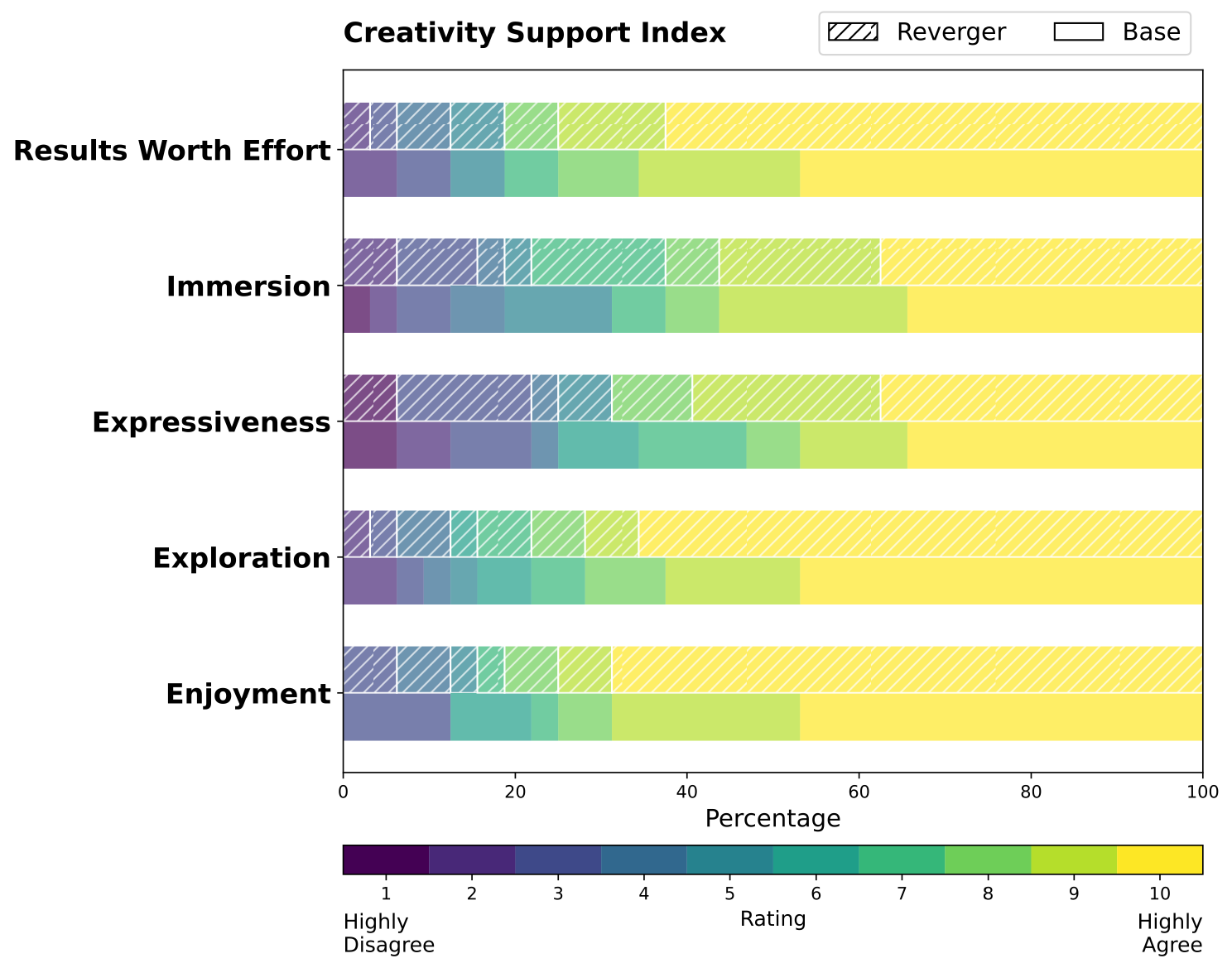}
        \caption{CSI Survey Results}
        \label{fig:csi}
    \end{subfigure}
    \hfill
    \begin{subfigure}[b]{0.48\textwidth}
        \centering
        \includegraphics[width=\textwidth]{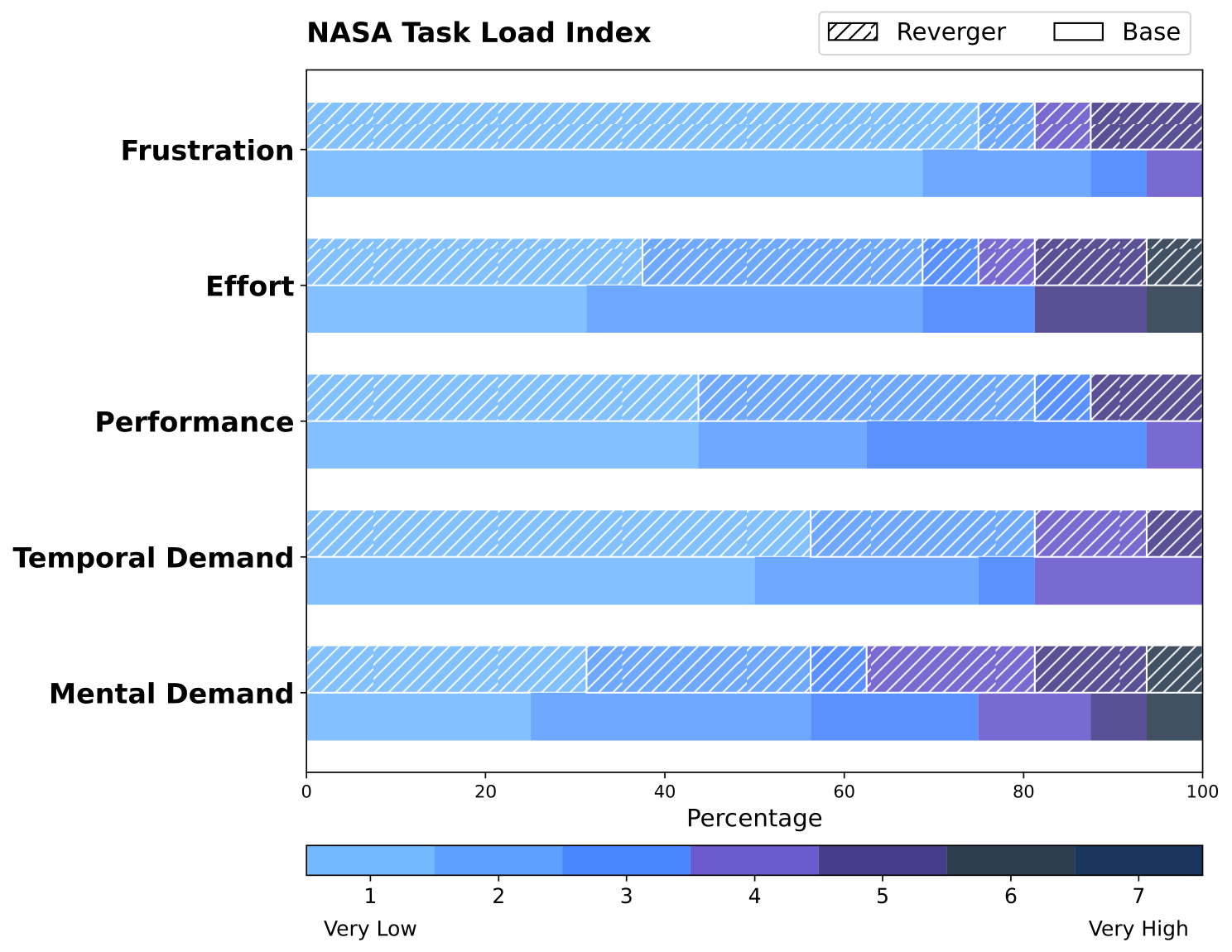}
        \caption{NASA-TLX Survey Results}
        \label{fig:tlx}
    \end{subfigure}
    \caption{Survey Results}
    \label{fig:survey_results}
\end{figure*}

\subsubsection{Post-study Interviews (25 mins)}
\label{method:interview}
After completing both tasks (one with the baseline and another with \sys{}), we ran a post-study interview. We began by asking participants to compare the two tools in their own words based on their overall impression. Next, we questioned how divergence and convergence supporting features of \sys{} assisted them in their tasks (RQ1) and how they would leverage those features into their future workflow (RQ2). For the questions about RQ1, we pointed out specific features one by one (\eg~\faSearch~(D.1),~\Rewind~(C.2)) to guide participants' understanding of our questions. We also asked questions regarding potential concerns and any suggestions for improvement. To conclude the study, we let participants ask any questions and share final comments.

\subsection{Measures}

We had both quantitative (survey) and qualitative (interviews and interaction logs) measures to understand how participants interact with our tool and the baseline, and how they perceive the interactions. We collected NASA-TLX and CSI survey responses as mentioned in~\autoref{method:task}. The CSI questions focused on enjoyment, exploration, expressiveness, immersion, and whether the results of tool usage were worth the effort. We excluded questions about \texttt{collaboration} for a few reasons. First, \sys{} was not intended to support \texttt{collaboration} with other creators. Second, our tool features do not provide agentic interaction that users might \texttt{collaborate} with AI agents to achieve tasks. Lastly, comparable CST research, like Luminate~\cite{suh2024luminate}, did not measure \texttt{collaboration} score. Then, we collected users' interaction log data to investigate their behaviors, including manually added directions, diverged directions, depth of exploration, converged multiple directions, generated mutant stories, and final outcomes. To secure the context and enable integrated interpretations with other data, we also recorded audio and shared screens during the study.

\section{Findings}

We first summarize our quantitative findings from the post-session survey (CSI in~\autoref{fig:csi} and NASA-TLX in~\autoref{fig:tlx}) (\autoref{finding:quant}).
Then we present how interaction logs and participants' interviews substantiate and further expand these quantitative findings, focusing on three key areas: i) the impact of \sys{} features supporting divergence-convergence workflow in the story ideation practices, ii) participants' envisioned integration of our tool into their workflows, and iii) any concerns about our tool.

\subsection{Summary of Survey Results}
\label{finding:quant}

We conducted the Wilcoxon Signed-Rank test to compare the baseline and our system across five abstract directions of both CSI and NASA-TLX survey results, because of the ordinal nature of the Likert scale data used in both surveys~\cite{choi2024creative}. For the CSI, we found significant differences between the two tools in providing \texttt{enjoyment} ($z = -4.56$, $p = .02$), \texttt{exploration} ($z = -4.56$, $p = .038$) and \texttt{results worth effort} ($z = -4.82$, $p = .003$). This suggests that our interface design facilitates the recursive exploration of abstract directions and variations through its divergence and convergence features (\texttt{exploration}). This improved exploration might enable users to have better control and increased chances to discover their desired outcomes (\texttt{results worth effort}) with more enjoyable experiences (\texttt{enjoyment}). However, there were no significant differences for \texttt{expressiveness} ($z = -4.04$, $p = .293$), and \texttt{immersion} ($z = -3.58$, $p = .847$).

The NASA-TLX results showed no significant differences between the baseline and our tool in all five directions: \texttt{mental demand} ($z = -2.95$, $p = .609$), \texttt{temporal demand} ($z = -3.31$, $p = .705 $), \texttt{performance} ($z = -3.13$, $p = 1.0$), \texttt{effort} ($z = -2.22$, $p = .794$), and \texttt{frustration} ($z = -3.43$, $p = .194$). At least, for our sample, the TLX scores do not present significant differences between the two systems. In summary, our participants perceived that our tool could facilitate exploring more variations and crafting more satisfactory outcomes. 
However, participants did not perceive cognitive loads to be significantly different between conditions.

\begin{table*}
\caption{Log data of \textit{Reverger} and the baseline. The \textit{Level of Depth} columns show the total number of directions revealed to participants, while the \textit{\# of Converged Directions} columns present the total number of attempts to converge multiple directions for generating tailored text variations. Numbers in the (base) columns show the baseline log data: the 2nd level of depth and variations with a single direction. *A single-direction choice. **While P12 manually added seven abstract directions, they had only two distinct ones. \dag~The number of directions that users added and specific examples when using \sys{}.}
\label{tab:log}
\centering
\resizebox{\textwidth}{!}{%
\begin{tabular}{l l l l l l l l l l l l l l l l l }
\toprule
    & \multicolumn{7}{c}{\textit{Level of Depth}} & \multicolumn{8}{c}{\textit{\# of Converged Directions}} & \\ \cmidrule(lr){2-8} \cmidrule(lr){9-16}
    
    & (base) & 2nd & 3rd & 4th & 5th & 6th & 7th & (base) & one* & two & three & four & five & six & seven & Added directions\dag \\
    \midrule
    
    P1 & 16 & 54 & 4 & 4 & - & - & - & 3 & - & - & 2 & 1 & 1 & - & - & 2: family, intentions \\ 
    P2 & 32 & 16 & 16 & 12 & 8 & 4 & - & 7 & 1 & 3 & - & - & - & - & - & - \\ 
    P3 & 32 & 30 & 8 & 14 & - & - & - & 6 & 2 & 2 & 1 & - & - & - & - & - \\ 
    P4 & 12 & 54 & 4 & - & - & - & - & 9 & - & - & 2 & - & - & - & 1 & - \\ 
    
    \hdashline
    P5 & 40 & 51 & 8 & - & - & - & - & 12 & - & 8 & - & - & - & - & - & 2: resolution, setting \\ 
    P6 & 73 & 54 & 17 & 4 & - & - & - & 14 & 8 & 6 & 3 & - & - & - & - & 4: hopeful, fantasy, relief \\ 
    P7 & 20 & 18 & - & - & - & - & - & 23 & 12 & - & 5 & - & - & - & - & 1: imagination \\ 
    P8 & 32 & 48 & 43 & 24 & 8 & 8 & - & 24 & 11 & 1 & - & - & - & - & - & - \\ 
    
    \hdashline
    P9 & 8 & 4 & - & - & - & - & - & 3 & 2 & 2 & - & - & - & - & - & 2: mysterious, sarcasm \\ 
    P10 & 35 & 16 & - & - & - & - & - & 31 & 13 & 3 & - & - & - & - & - & 2: excitement, dreamy \\ 
    P11 & 29 & 36 & 12 & - & - & - & - & 6 & - & 2 & 7 & 1 & - & 2 & - & 1: fun \\ 
    P12 & 44 & 61 & 16 & 8 & 12 & 4 & 4 & 7 & - & 4 & 1 & - & - & - & - & 7**: sinister, first-person \\ 
    
    \hdashline
    P13 & 24 & 54 & - & - & - & - & - & 5 & - & 1 & 1 & 1 & - & - & - & - \\ 
    P14 & 15 & 25 & 4 & - & - & - & - & 4 & - & 7 & 2 & - & - & - & - & 1: clothes \\ 
    P15 & 55 & 48 & 20 & 4 & - & - & - & 10 & 2 & 6 & 3 & - & - & - & - & 2: tragic, action \\ 
    P16 & 28 & 54 & 27 & 8 & 4 & - & - & 14 & - & 3 & 3 & 3 & - & - & - & 12: area 51, The Gladiator \\ 
\midrule
    Total & (495) & 623 & 179 & 78 & 32 & 16 & 4 & (178) & 51 & 48 & 30 & 6 & 1 & 2 & 1 \\
\bottomrule
\end{tabular}
}
\end{table*}

\subsection{Impact of Divergence and Convergence Features in Story Ideation Practices (RQ1)}
\label{findings:rq1}

All participants recognized that \sys{} offers more options for users to control. It supports participants' recursive exploration of high-level directions to modify the original story, and convergence of multiple directions to generate synthesized artifacts. Also, all participants, except P12, preferred \sys{} over the baseline. P12 wanted fewer options (see~\autoref{finding:div} for details). In this section, we present both interaction log data and interview quotes to explain how divergence and convergence supporting features of \sys{} specifically facilitate users' story ideation workflow, highlighting the tool's impact on creative processes and outcomes.

\subsubsection{Divergence Features Expand the Boundary of Creativity}
\label{finding:div}

Participants (\eg P1, P5, P7, P8, P11, P13, P14) reported that divergence features (\faSearch~(D.1) and \faPlusCircle~(D.2)) expanded their creative boundaries. With recursive exploration, of course, users could explore deeper sub-directions. Surprisingly, users also explored more 2nd level sub-directions with \sys{} than with baseline (see~\autoref{tab:log}). The affordances to explore deeper directions might have facilitated the users to explore more on the surface level as well. Specifically, participants explained that our system generated unexpected directions that participants had not previously considered at all. This divergent exploration and discovery stimulated participants' imaginations, leading to more diverse ideas. For example,

\begin{quote}
P5: Things of things that I wouldn't normally gravitate towards. so it could possibly give me new ideas.
\end{quote}
\begin{quote}
P8: Manic; I wouldn't have thought about it if I hadn't been able to play with those `emotions' [recursively].
\end{quote}
\begin{quote}
P11: It was helpful to me, honestly, because it gives you keywords that you may not even think about.
\end{quote}

Moreover, the expanded exploration allowed participants to identify and refine specific aspects they wanted to change. Many participants initially had only vague intentions, such as `making it emotional' However, the \faPlusCircle~(D.2) enabled them to recursively explore more specific options (see~\autoref{tab:recursive}). For example, P11 particularly liked that `emotion' could be further refined into `happy, sad, anger', and `sad' could be further broken down into `depressed, loss, melancholic'. This feature allowed participants to carefully reconsider and specify their desired directions. Log data revealed that our system provided an additional 309 directions across the third to seventh levels of depth (see~\autoref{tab:log}), a feature not available in the baseline system without the \faPlusCircle~(D.2) (see~\autoref{study:baseline} for the baseline setup).

\begin{table*}
\caption{Hierarchical exploration of abstract directions to modify the original story across three levels. Bold text in the second level indicates parent directions for the third-level entries.}
\label{tab:recursive}
\centering
\resizebox{\textwidth}{!}{%
\begin{tabular}{l l l l l}
\toprule
    & 1st & 2nd & 3rd \\
    \midrule
    P1 & Characters & species, identity, profession, \textbf{relationships} & Family, Romance, Friendship, Rivalry \\
    P4 & Theme & Romance, Adventure, Mystery, \textbf{Love} & Romantic, Platonic, Familial, Unconditional \\
    P8 & Plot & Conflict, \textbf{Climax}, Resolution, Foreshadowing & Culmination, Confrontation, Tension, Uncertainty \\
    P12 & Settings & \textbf{Location}, Era, Landscape, Environment & Terrain, Scenery, Climate, Topology \\ 
    P14 & Romance & Passion, \textbf{Intimacy}, Commitment, Affection & Physical, Emotional, Spiritual, Mental \\
    P16 & Military & Weapons, \textbf{Codes}, Training, Ranks & Societal, Cultural, Religious, Linguistics \\
\bottomrule
\end{tabular}
}
\end{table*}

While divergence supporting features are generally perceived as enhancing the diversity of options to explore, some participants also raised a potential drawback. Specifically, P12 was concerned about being overly immersed in many options with playful dynamics.

\begin{quote}
P12: I could focus more on the task with the second one [the baseline]. I liked that I could do more with the 1st one [\textit{Reverger}]. But for me, it was a little bit like, `\textit{Oh, what's this one? And what happens if I do this? And oh, that sounds interesting. Oh, but this sounds interesting, too.}' 
\end{quote}

Immersive experiences with diverse options may seem less efficient from a task-oriented perspective, as users might spend excessive time exploring possibilities rather than completing story ideation tasks. However, in autotelic creative contexts, where the process is valued alongside the end product, these experiences can potentially enhance user engagement and satisfaction (\eg casual creators~\cite{CasualCreators}, reflective creators~\cite{ReflectiveCreators}).

\subsubsection{Convergence Features Enable Granular Instantiation of Ideas}
\label{finding:granular}
We found that convergence supporting features, specifically \faCheckSquare~(C.1) and \Rewind~(C.2), improved efficiency and precision in achieving desired outcomes. First, participants (\eg P1, P2, P4, P7, P8, P11, P13, P14) reported that \textit{Reverger} reduced the number of iterations needed to discover satisfactory variations. User interaction logs showed that participants generated 127 synthesized variations across variations with multiple directions using our system, compared to 178 mutants with only a single direction (see~\autoref{tab:log}). This indicates that \sys{} enabled participants to discover their desired outcomes in fewer iterations while achieving higher satisfaction levels (\texttt{\textbf{results worth effort}} in~\autoref{finding:quant}).

\begin{quote}
P2: It [\textit{Reverger}] was able to let me select more categories of what I wanted to see, and it gave me the answer I wanted to see right away, whereas in the second one [baseline], I was limited in how I could select certain ones, and that never gave me the right answer that I really wanted.
\end{quote}

\begin{quote}
P13: [In baseline] I did not like having to apply three different things at three separate times. However, I can easily do one at a time in the other system [\textit{Reverger}], but I will also have the opportunity to combine them together to see what that does all together, instead of reading three different instances.    
\end{quote}

Furthermore, these convergent features empowered users to achieve their precise desired outcomes by generating variations across multiple selected directions simultaneously. This level of control resulted in higher-quality outputs compared to the baseline system, which produced less satisfactory results.

\begin{quote}
P2: I was limited in how I could select certain ones [with the baseline], and that [baseline] never gave me the right answer that I really wanted. Even if it [baseline] did give you the right answer, I wasn't truly impressed with it.
\end{quote}
\begin{quote}
P8: This [\textit{Reverger}] gave me almost exactly what I was looking for when I selected those two criteria.
\end{quote}

While many participants found value in the multiple selection feature, others, like P10, felt it was not particularly necessary for their needs. P10 explained that she was able to achieve her desired results using only a single-direction choice.

\begin{quote}
P10: It was already close [The generation with a single direction is already close to the desirable outcome], so I didn't need it [multiple selections] to change five different things at one time.
\end{quote}

\begin{figure*}
  \includegraphics[width=\textwidth]{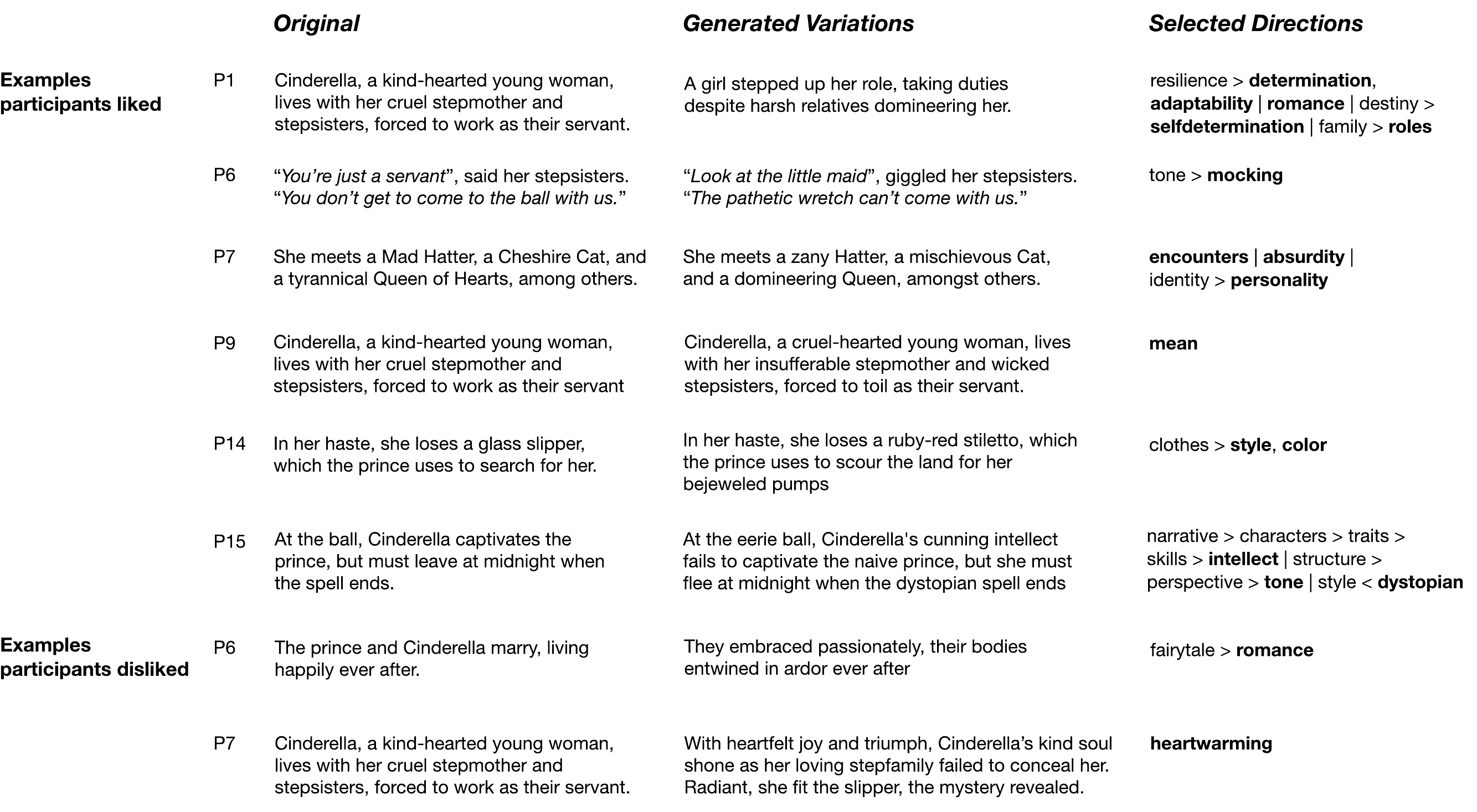}
  \caption{Generated variations by participants. In the \textit{Selected Directions} column, we highlighted in bold the directions chosen by each participant. For instance, P7 selected three directions: \textbf{encounters}, \textbf{absurdity}, and \textbf{personality}.}
  \label{fig:examples}
\end{figure*}

Indeed, participants' reactions to outcomes varied depending on the amount of changes they wanted to make. Some participants (\eg P6, P7) liked outcomes with minimal alterations, such as changes in tone or with more detailed descriptions, while maintaining other elements of the original story. For example, P6 was impressed by the \textbf{mocking} tone, which was her choice, in the refined dialogue ``\textit{Look at the little maid}'', transformed from ``\textit{You're just a servant}'' (see~\textit{examples participants liked} in~\autoref{fig:examples}). She typed dialogues directly in the draft, as this has always been challenging for her. To her surprise, the resulting quality was exceptionally good. However, these same participants (P6 and P7) strongly objected to generated variations where changes were more significant (see~\textit{examples participants disliked} in~\autoref{fig:examples}). They even considered those examples as incorrect and misinterpretation.

\begin{quote}
P6: It is not even what the sentence is talking about. It's misinterpreted what's being said here.
\end{quote}

Conversely, other participants, such as P1 and P9, enjoyed making significant modifications by selecting multiple directions, even to core aspects of the original stories. For instance, both P1 and P9 were pleased to revise Cinderella as a rebellious and mean girl, departing from her traditionally modest and humble portrayal. Also, P15 enjoyed a twist where Cinderella fails to attract the prince, unlike in the original story (see detailed examples in~\autoref{fig:examples}). These varying preferences suggest that users might want to calibrate the range of generated variations in these types of AI-powered co-creative systems, which we discuss further in~\autoref{dis:calibrate}.

\subsection{Participants' Envisioned Integration of \sys{} into Creative Writing (RQ2)}
\label{findings:rq2}

Participants envisioned various ways of integrating this tool into their creative writing process, ranging from i) minimal changes that are not dependent on other parts (\autoref{rq2:minimal}) to ii) major brainstorming with significant alterations (\autoref{rq2:big}). This range of uses can be viewed as a spectrum, with individual writers falling somewhere along it.

\subsubsection{Utilizing to Refine and Enrich Targeted Passages}
\label{rq2:minimal}

Some participants (\eg P3, P6, P7, P9, P10) were more inclined to use the tool for enhancements that enriched specific passages without affecting the broader narrative. These changes included several stylistic variations such as adjusting tone, adding more descriptive words/phrases, and refining dialogue details. For example, 

\begin{quote}
P3: I can't really think of a joke for a character to say in a certain context. Then I highlight that part to know the end of the joke, to see what it comes up with. That's when I think this [\textit{Reverger}] would be most useful at that little bite-size level. But I would not put a whole chapter.
\end{quote}
\begin{quote}
P6: I thought the dialogue was really good. This could be really useful for a lot of writers because writing dialogue that sounds emotive and more flowing and natural is definitely something a lot of writers have trouble with.
\end{quote}

Both P3 and P6 were reluctant to get help with the ideation and creative process from our system. This reluctance appears in the log data (see~\autoref{fig:examples}). For instance, P6 appreciated the tool's ability to enhance tone and effectively communicate situations and emotions. However, she was averse to generated variations that strayed too far from her intended narrative direction.

\subsubsection{Utilizing to Discover and Instantiate New Ideas}
\label{rq2:big}

Some participants (\eg P1, P7, P8, P11, P14, P16) were more intrigued by the system's ability to generate and explore diverse ideas to make even content-related changes. Specifically, some (\eg P1, P7, P16) wanted to use \textit{Reverger} to generate various ideas to instantiate their vague ideas into more concrete artifacts. For example,

\begin{quote}
P7: It [story] doesn't always come out the way you really imagined. So, if you can just put a rough draft, this [\textit{Reverger}] can help you get exactly what you're trying to find. It can fine-tune it [narrative] for you.
\end{quote}
\begin{quote}
P16: There are times that I haven't followed through with writing something that I had a really cool idea. If I were to put a gist of my idea, I'd use this tool [\textit{Reverger}] to layer and maybe make it more robust. It would be easier to write the next sentence.
\end{quote}

Other participants (\eg P8, P11, P14) envisioned leveraging our system to discover new narrative directions, even if they significantly altered their overall stories. For example,

\begin{quote}
P8: For genre literature where you're taking something classic, this would be amazing to just brainstorm and give you a sense of direction that might work with different audiences.    
\end{quote}
\begin{quote}
P14: At the touch of the key, it can generate lots of ideas. It'd be also interesting to see what it could do in terms of plot ideas. I can pick from the menu on the right what to do with it, and direct it.
\end{quote}

\subsection{Concerns and Ownership}
\label{findings:concerns}

Participants expressed two main concerns regarding \sys{}: inconsistency issues (\autoref{concern:consistency}) and over-reliance on AI (\autoref{concern:overreliance}). Additionally, many participants showed a surprisingly strong sense of ownership over their outcomes, despite the support of the LLM (\autoref{concern:ownership}). 

\subsubsection{Localized Exploration Entails Inconsistency of Narrative}
\label{concern:consistency}
The system's affordance to highlight and revise specific parts is beneficial, as it allows writers to preserve the other parts unchanged (as mentioned by P11). However, several participants (\eg P3, P6, P15) noted that editing only highlighted draft passages could lead to narrative inconsistencies. For example, modifying certain character aspects may require more comprehensive changes throughout the entire story, not just in the highlighted sections. 

\begin{quote}
P3: You can't just make a character more `resilient’ by changing the phrase of a sentence. Maybe if you did the whole story, it may be resilient.
\end{quote}
\begin{quote}
P15: [after updating a part] I just needed to adjust the flow. Not just the tone. Yeah, to make the story consistent.
\end{quote}

Therefore, AI tools should facilitate maintaining the consistency of artifacts~\cite{UnmetNeeds}. We discuss several potential approaches to address this in~\autoref{dis:consistency}.

\subsubsection{Potential Overreliance on AI}
\label{concern:overreliance}

One participant (P6) expressed concern about potential over-reliance on \sys{}. She acknowledged its clear benefits: supporting brainstorming, ease of use, and reducing the cognitive load for creative writing. Additionally, the tool's immersive and enjoyable nature (\texttt{\textbf{enjoyment}} in~\autoref{finding:quant}) could attract many users, potentially resulting in excessive dependence on this system (a concern also shared by P12 in~\autoref{findings:rq1})~\cite{ComplementaryCognitiveArtifacts}.

\begin{quote}
P6: I can definitely imagine a lot of writers sort of really relying, overly reliant on this. Because you could have a really basic story and essentially get the system to rewrite for you.
\end{quote}

This over-reliance could raise several additional issues. For example, P3 and P14 were concerned that extensive use of this tool could lead to potential copyright infringement, as AI-generated suggestions might be retrieved or recombined from existing online sources. Also, they noted the unclear distinction between writers' input and the AI output, making plagiarism concerns more evident. P6 suggested that the AI tool should present a clear breakdown of content portions to clarify the extent of AI assistance like a word count feature (\eg~\cite{Aillude}). This approach could help determine responsibility and rights while mitigating the risk of over-reliance on AI.

\subsubsection{Ownership}
\label{concern:ownership}

Interestingly, many participants in our study felt a strong sense of ownership over their ideation outcomes, attributing this to the time and effort they invested in the process. This finding contradicts previous studies that reported people losing a sense of ownership when using LLMs (\eg~\cite{kadoma2024role, dhillon2024shaping}). 

\begin{quote}
P14: I think it [ownership] would have to be strong in the sense that if you have the option of keeping it or omitting it. So if you're building it piece by piece, it, in the end, reflects what you've put in. It's the end result of your choices.
\end{quote}

Participants emphasized that while AI provided various options for story ideation, all final choices and decisions were made solely by themselves. Additionally, many participants manually added unique directions to their work, which they believed would be difficult for others to replicate exactly (\eg P16). These factors further reinforced their sense of ownership over the final product.

\begin{quote}
P16: I don't know how egotistical it is, but if you're applying multiple filters, I feel like it's a hundred percent what I made. The likelihood statistically, of someone else selecting the same sentence, applying the exact same outputs, considering the search bars, is very small. So I'm gonna say ownership is mine.
\end{quote}

However, this sentiment was not universal. A few participants (P3, P5, P7, and P15) still did not feel a strong sense of ownership over their final outcomes. For example, P3 compared the experience to `buying a cupcake from a convenience store', stating that he would never claim such a purchase as his own work.

\section{Discussion and Future Directions}

\subsection{Scaffolding Divergence and Convergence for Creative Ideation}
\label{dis:proximal}

We designed the interface to provide space for users to infuse and refine their intentions throughout these iterative processes, aiming to give users more controllable interactions with AI. While participants showed promising responses to our approach, how they envisioned the utility of our tool in creative practice varied (\autoref{findings:rq2}). Reflecting on these diverse responses, we discuss additional control features users desire and how we could calibrate these controls to improve iterative cycles of divergence and convergence in creative tasks. Then, we discuss the applicability of supporting recursive divergence and convergence in other domains.

\subsubsection{Calibrating User Control of LLM Output Diversity}
\label{dis:calibrate}

The current interface design of \sys{} does not allow users to calibrate the diversity of abstract ideas generated by the LLM. In our creative writing context, we observed a spectrum of user needs to control the diversity scope of LLM outputs. Some participants preferred LLM ideation output that preserves the existing narrative while making other changes, such as converting descriptive passages into dialogue or refining specific expressions. Others wanted to explore fundamentally different narrative directions, such as alternating character personalities and introducing new plot points. Although our design enables users to specify which parts of a draft may or may not be modified, it lacks direct control for setting constraints or diversity parameters within those highlighted areas. Implementing more granular control over the diversity scope of LLM outputs would enhance the interaction paradigm.

Future studies could pursue both model-centric and interaction-centric approaches to calibrating user control of the diversity of LLM output. AI models can learn and adapt through iterative processes based on user interactions~\cite{gao2024aligning}. For instance, if users consistently reject suggestions aligned with the existing narrative, a model could evolve to generate more diverse alternatives. Such systems can also leverage ``non-selected'' options as negative examples to give negative weights to LLM outputs~\cite{chris2017deep}. Although these model-centric approaches would be promising, we strongly encourage the UIST community to consider interaction-centric approaches through a novel interface design.
For example, metaphoric interactions similar to TextoShop~\cite{masson2025textoshop} could be one way to control the diversity scope of LLM output. We envision interfaces where users might `extract' a language tone using a metaphorical `color picker' from an original draft, then `drop' this extracted sample (tone) to a palette~\faPalette~to explore stylistic variations while preserving core narrative elements. Another direction might be to leverage physical interaction techniques to enable diversity calibration (\eg~\cite{lin_semanticpinching_2025}). Drawing from familiar gestural interactions like pinch-to-zoom in image manipulation, where users naturally understand that pinching controls granularity of detail, similar gestures could calibrate LLM output diversity during creative ideation.
Such controllability through interaction could provide even more fine-grained control over the diversity of LLM output, which may improve creators' sense of ownership (we discuss ownership further in~\autoref{dis:agencyhai}).

\subsubsection{Handling Consistency in Divergence and Convergence}
\label{dis:consistency}

Our tool allows users to work on specific parts of the original artifact, offering flexibility in creative processes. However, we found that this localized editing approach can lead to consistency issues throughout the artifact (\autoref{concern:consistency}). In creative writing contexts, when authors make changes to a character, these alterations can create chain reactions throughout the story~\cite{mani2013computational}. Therefore, an AI-powered tool should be designed to address these consistency issues. For example, AI tools for supporting divergence and convergence iteration should be capable of identifying potential inconsistencies resulting from localized edits~\cite{LooseEnds} and offer suggestions to preserve consistency throughout the entire narrative. In addition, the tool should provide options for users to decide how they want to handle these subordinate consequences. By empowering users to make informed choices about the scope and impact of their revisions, AI-powered tools can effectively improve authorial control and creative satisfaction. These consistency issues would not be limited to creative writing, but could be applied to other creative domains.

\subsubsection{Supporting Divergence and Convergence in Other Domains}

Supporting recursive divergence-convergence processes can be extended to various creative domains beyond writing, such as music composition and image generation. For example, prompt-based image generation tools (\eg Midjourney~\cite{midjourney2025}, Stable Diffusion~\cite{rombach2022high}) typically produce concrete artifacts. However, users often desire to refine specific elements (\eg~\cite{zhang2024transparent, chung2023promptpaint}) within these generated images. To address this need, these tools could implement a more sophisticated editing process. Users could highlight particular areas of an image, and the AI model could then generate high-level directions to modify the given images. This approach would offer users a wide range of options, from subtle variations to dramatically different alternatives with their own control. Furthermore, users could select and integrate multiple directions to edit images~\cite{choi2024creative}, providing greater human control in creative processes. Similarly, user interface (UI) generation tools (\eg~\cite{wu2024framekit}) could benefit from adopting this approach. As UI design is also inherently iterative~\cite{wu2019ui}, incorporating divergence-convergence support could improve the productivity and efficiency of the design process~\cite{dow2011parallel}. Designers could explore various layout options, color schemes, or interaction patterns, then converge on preferred elements to refine further.

\subsection{Towards Preserving Human Agency within Human-AI Interaction}
\label{dis:agencyhai}

Many participants felt a strong sense of ownership over their creative output, which could be attributed to our interface design enabling improved controllability over LLM outputs. This led us to consider the balance between productivity and ownership in AI-assisted creative work. The relationship between control, ownership, and creative satisfaction emerged as an interesting finding in our study. Finally, we discuss interface design philosophies that better align AI output with human intentions, with a particular focus on preserving human agency within the human-AI interaction.

\subsubsection{Balancing between Productivity and the Sense of Ownership}
\label{dis:auth}

Our study gives us hints to extend the longstanding discussion about how to design a human-AI system to balance between automation and control~\cite{ben1997direct, jeff2019agency, brian2019ask}. We demonstrated that an interface offering a structured series of interactions
enabled users to achieve higher-quality exploration of diverse ideas and more satisfying outcomes in creative story ideation. Several participants (\eg P2, P13, P14) reported a lower perceived cost (\ie time and effort) and cognitive burden, while achieving more satisfactory outcomes with \sys{} compared to a conventional creative process without AI. Interestingly, many participants felt a strong sense of ownership over their artifacts despite investing less cognitive effort, which contradicts the conventional belief that the sense of ownership is often cultivated through investment of time and effort (cognitive burden)~\cite{rochberghalton1984object, pierce2001to}. This finding contrasts with previous research suggesting that AI-powered CSTs compromise users' sense of ownership~\cite{kadoma2024role, dhillon2024shaping}. This enhanced sense of ownership might be attributed to the fine-grained control offered by our interface design.

While existing AI tools (\eg ChatGPT) often generate entire artifacts from scratch (automation), potentially diminishing users' authorial control~\cite{kim2024authors, draxler2024ghost}, our tool allows users to selectively identify and modify particular elements (control) through recursive exploration and convergence features. We deliberately designed the system to allow users to make their own judgments about LLM-generated high-level directions and to infuse their personal preferences into the outcome. This design seems to encourage users to invest more careful consideration and effort into decision-making, which may contribute to their increased sense of ownership despite the reduced overall cost. Hence, it strikes a balance between AI assistance and human agency.
Future research should investigate further mechanisms for balancing or even fostering productivity without compromising ownership with the help of LLMs. Some relevant research questions worth investigating would be: What interactions effectively impart a feeling of control? What facets of ownership and control should be prioritized in digital interfaces to enhance creative experiences?

\subsubsection{Interface Design for Human-AI Alignment}

While the machine learning community focuses on improving models to address human-AI alignment issues (\eg~aligning AI assessments with human judgments \cite{shaikh2025aligning}), we should consider which domains benefit most from these model-centric approaches. For tasks requiring high precision with an objective ground truth, such as text classification~\cite{thomas-etal-2024-never} and predictive analysis~\cite{jiang2025reasoningenhanced}, the value of model-centric alignment approaches is 
clear. However, this might not be the case in creative domains that involve iterative divergence and convergence processes, where individual uniqueness and subjective interpretation are valued (\eg creative arts, science activities). In this context, we should be cautious about delegating human judgment to AI systems. Human-AI collaboration should be interactive and iterative to let users align outcomes with their intentions through appropriate control mechanisms. We therefore invite the UIST community to consider that designing interfaces for expressing meaningful human intention might be a complementary approach to addressing human-AI alignment problems.

Consider AI for scientific research as an illustrative case. AI can enhance numerous aspects of the research pipeline, including generating research topics~\cite{pu2025ideasynth, liu2024how, si2025can}, simulating scenarios~\cite{park2022social}, and analyzing existing literature~\cite{asai2024openscholar}. However, there is an increasing trend toward automating significant portions of the scientific process (\eg~OpenAI DeepResearch~\cite{openai2025introducing}). We argue that the HCI community should steer this research toward developing interfaces that preserve human control in scientific research, rather than fully delegating accountability to generative AI. As scientists, we have the responsibility to carefully consider and proactively direct these technologies in ways that benefit society~\cite{bird2014socially}. This requires critical examination of whether diminishing human involvement in research activities truly serves the advancement of human welfare and knowledge production. This calls for the development of interfaces that preserve the essential human elements of scientific inquiry while leveraging AI capabilities as supportive tools.

\subsection{Limitations}
\label{dis:limit}

Our study has several limitations regarding technical dependencies and the nature of the lab study design. The generated variations and suggestions of \sys{} heavily rely on Anthropic's Claude model. Employing different LLMs could change the overall user experience and the quality of story ideation outcomes. In addition, the use of well-known stories (Cinderella and Alice in Wonderland) may have influenced the user experience and outcomes, as these stories were likely included in the training data for the LLMs. Future studies could investigate how well-known stories could potentially constrain or promote the diversity of LLM-generated suggestions compared to individual creative writers' novel artifacts.

Our study design could bring another limitation. Our lab study with 16 participants using pre-selected familiar stories rather than personal writing may not reflect authentic usage contexts. Also, the participants' evaluation (\eg survey and interviews) was based on a brief 10-minute exploration for each session. That being said, the `novelty effect' might dominate in our findings to an extent that they might not with longer-term use~\cite{EvaluatingCSTs}. A longitudinal study could provide more comprehensive insights. Therefore, in future studies, we should run a longer-form study where writers use \sys{} to work on their own artifacts over an extended period (\eg~\cite{NotJustNovelty, jacobs_supporting_2017}). This approach would provide more in-depth insights into the practical implications of our interface design and its impact on the `real-world' writing process. Lastly, all participants were hired from a freelancing platform. Therefore, our data collected from those freelancers may not fully represent the potential reactions from other creative writers in the wild.

\section{Conclusion}

We designed and implemented \sys{}, which scaffolds recursive divergence and convergence in creative story ideation with the help of LLMs. With our tool, users can highlight any parts of the original story and explore multiple high-level directions to modify the highlighted part. Then they can recursively explore sub-directions of each high-level direction for more granular exploration and refinement. \sys{} enables the selective collection of multiple directions to tailor generated variations. Through this process, our tool lets users explore more granular examples and refine their intention to modify story concepts. By providing tangible variations based on user-selected directions, our tool enables users to converge their ideas to discover satisfactory outcomes for ideation. We conducted a user study with 16 participants, ranging from novice to expert writers, to evaluate \sys{}'s effectiveness and impact on the creative story ideation workflow. Participants rated our tool as providing significantly better exploration and immersive experiences, with outcomes deemed more effort-worthy than the baseline. Our log data and participant interviews further support that \sys{} enabled users to explore more divergent high-level directions and discover a satisfactory outcome with fewer generations of variation ideas. We reflect on supporting divergence-convergence processes with AI for creative ideation and designing interfaces for human-AI alignment. We hope this work can offer insight for designing interfaces of AI tools supporting divergence and convergence processes with human agency in various creative domains.




\bibliographystyle{ACM-Reference-Format}
\bibliography{references}

\end{document}